\documentclass[floatfix,
aps,prb,
 amsmath,amssymb,
 reprint,%
]{revtex4-1}

\usepackage{graphicx}
\usepackage{dcolumn}
\usepackage{bm}
\usepackage{xcolor}
\usepackage{soul}

\begin{document}

\preprint{AIP/123-QED}

\title{Selective band engineering of Bi/Si(111) by boron segregation}

\author{E. Barre}
\email{etienne.barre@espci.fr}
\affiliation{ 
 Laboratoire de Physique et d’Étude des Matériaux (LPEM), ESPCI Paris, PSL University, CNRS UMR8213, Sorbonne University, 75005 Paris, France}%
\author{J. Villalobos-Castro}%
\affiliation{ 
Université Paris-Saclay, ONERA, CNRS, Laboratoire d'étude des microstructures (LEM), 92322, Châtillon, France}%
\author{T. Pierron}%
\affiliation{ 
 Laboratoire de Physique et d’Étude des Matériaux (LPEM), ESPCI Paris, PSL University, CNRS UMR8213, Sorbonne University, 75005 Paris, France}%
\author{S. Pons}
\affiliation{ 
 Laboratoire de Physique et d’Étude des Matériaux (LPEM), ESPCI Paris, PSL University, CNRS UMR8213, Sorbonne University, 75005 Paris, France}%
\author{D. Roditchev}
\affiliation{ 
 Laboratoire de Physique et d’Étude des Matériaux (LPEM), ESPCI Paris, PSL University, CNRS UMR8213, Sorbonne University, 75005 Paris, France}%
\author{L. Sponza}
\affiliation{ 
Université Paris-Saclay, ONERA, CNRS, Laboratoire d'étude des microstructures (LEM), 92322, Châtillon, France}%
\author{S. Vlaic}
\affiliation{ 
 Laboratoire de Physique et d’Étude des Matériaux (LPEM), ESPCI Paris, PSL University, CNRS UMR8213, Sorbonne University, 75005 Paris, France}%

\date{\today}

\begin{abstract}
Atomically thin layers of metals deposited on semiconductors display a variety of physical properties such as superconductivity, charge density waves, topological phases, strong spin-orbit (Rashba) splitting, among others. To access these exotic phases and induce new ones, it is necessary to control and tune the energies of the electronic states of those heterostructures. In this work we investigate the engineering of the band structure of the two-dimensional Bi/Si(111) $\beta$-phase using a modulation doping approach based on boron segregation at the surface. We demonstrate that the Bi-induced Rashba bands can be displaced in energy by up to 200 meV without altering their strong Rashba parameter. Importantly, while the Bi states shift upward, the underlying Si valence states remain essentially fixed, which rules out a simple band-bending scenario. Our density functional theory calculations reveal that the displacement originates from changes in the hybridization of Si states near the surface, induced by the presence of B atoms. This selective mechanism halves the distance of the Rashba-split states from the Fermi level, opening the way to their exploitation in transport and spintronic devices and highlighting the broader potential of modulation doping as a band-engineering strategy for two-dimensional metals on semiconductors.

\end{abstract}

\maketitle

\section{Introduction}

Since almost two decades, two-dimensional metallic layers (2D metals) on semiconductors are attracting a strong scientific interest, both from the fundamental and applied point of view, due to their exceptional electronic properties. These materials can host, among others, superconductivity\cite{Zhang_2010,Brun2014,Menard2017,Wolf2022,Baranov2022}, two-dimensional topological phases\cite{Hsu_2015,Gruznev2018}, or charge density waves\cite{Mascaraque1998,Tejeda2007,Tresca2021}. The emergence of these intriguing electronic properties arises from their crystallographic structure, their reduced dimensionality, but also from the underlying semiconductor which imposes the position of the Fermi level. Controlling the filling of the band structure can therefore be an extremely important factor not only to enable applications but also to disclose novel and exotic electronic phases. 

Recently, studies on the Sn/Si(111) system\cite{Ming2017,Ming_2017,Yi2018,Wu_2020,Ming_2023} have highlighted the potential of modulation doping as a means to achieve such control. In this approach, the free charge carriers, brought by the semiconductor dopants that reside in a different spatial region with respect to the 2D metal atoms\cite{Dingle_1978}, induce a modulation of the Fermi energy position. When applied to the Sn/Si(111) system, switching from an n-doped Si(111) substrate to a boron (B) p-doped Si(111) substrate led to the emergence of new structural phases at low temperatures\cite{Ming2017}. Under appropriate conditions, modulation doping even enabled the emergence of unconventional superconductivity\cite{Wu_2020,Ming_2023}. The origin of these modifications resides in the change of the Fermi level position in the band structure of the two-dimensional Sn layer. 

The potential application of this approach goes far beyond the case of Sn/Si(111). Its generalization to other 2D metals, or more precisely to systems where interesting electronic features are located below the Fermi level, could pave the way to the disclosure of novel quantum phases or to the design of appealing systems for device fabrication. Prototypical candidates that could benefit from such band engineering are those with strong Rashba effect located below the Fermi level\cite{Gierz2009,Zhu2013,Gruznev_2014}. These materials are, in theory, excellent candidates for spintronic devices since they present extremely high Rashba parameters\cite{Han2018}, provided that the Rashba-split bands can be brought to the Fermi level. 

A paradigmatic case is a single atomic layer of Bi on Si(111), which has been extensively studied over the past two decades. It exhibits two different phases, called $\alpha$- and $\beta$-phases, corresponding to a monomer phase with a coverage of 1/3 monolayer\cite{Mono} and a trimer phase with a coverage of 1 monolayer respectively\cite{Hsieh_2020,Tsukanov2019}. In both cases the surface reconstructs adopting a $(\sqrt{3}\times\sqrt{3})R30^{\circ}$ structure with respect to the Si(111) substrate, and both phases possess Rashba-split states at the M point of the reconstructed Brillouin zone\cite{Gierz2009,Frantzeskakis_2010,Sakamoto_2009,Berntsen_2018}. The $\beta$-phase, in particular, presents one of the strongest Rashba parameters ever measured in similar systems\cite{Gierz2009,Frantzeskakis_2010,Hsieh_2020}. However, these Rashba states consistently lie well below the Fermi level. Recent theoretical calculations have shown that the $\beta$-phase is semiconducting, with the Bi-induced Rashba states positioned approximately 0.3 eV below the apex of the Si(111) valence band\cite{Chi2021,Gruznev_2014}. While the Rashba-split bands at the M point have mainly Bi character, the Bi electronic structure is strongly hybridized with the Si substrate at the $\Gamma$ point, which therefore imposes their energy position. Previous experimental studies of the $\beta$-phase used n-doped Si(111) substrates\cite{Gierz2009,Frantzeskakis_2010,Sakamoto_2009,Berntsen_2018}, where the Rashba crossing point was consistently found around 600 meV below the Fermi level. 

In this work, we apply modulation doping to the $\beta$-phase of Bi/Si(111) in order to explore to which extent this technique can be used to provide the desired displacement of electronic states toward the Fermi level. By controlling boron segregation at the surface, we can qualitatively adjust the local p-doping level, which we characterize by LEED. ARPES measurements reveal a systematic upward shift of the Bi Rashba bands as a function of boron segregation. With the help of density functional theory (DFT) we demonstrate that this shift does not originate from conventional band bending, but instead from a selective modification of the hybridization of Si states close to the surface.

\section{Material and Methods}

ARPES experiments were performed in an ultra-high vacuum (UHV) chamber (base pressure $\approx 1\times 10^{-10} \,\text{mbar}$) equipped with a SPECS Phoibos 225 analyzer, at $80 \text{K}$. He I ($h\nu = 21.218 \, \text{eV}$) UV radiation was produced using a SPECS UVLS source with p-polarized light. A custom-built 6-axis goniometer allowed exploration of reciprocal space. The samples were prepared \textit{in situ} from a p-doped (B, $ \rho=0.001-0.005 \,\Omega \cdot \text{cm}$) Si(111) substrate, which was cleaned through multiple cycles of Ar$^+$ ion sputtering at 1 keV, followed by ten seconds of annealing at 1320 K, until a clean $7\times7$ surface reconstruction appeared in LEED. Boron was segregated to the surface by annealing the substrate at 1400 K for ten minutes, confirmed by the appearance of a sharp $(\sqrt{3}\times\sqrt{3})R30^{\circ}$ pattern in LEED. For intermediate samples, the boron quantity was qualitatively controlled using LEED after sputtering and 1320 K annealing cycles. Bi was then deposited using a molybdenum crucible filled with high-purity (\textgreater 99.999\%) Bi pellets, at a rate of $0.28 \,\text{ML/min}$. All depositions were performed while maintaining the substrate at $\sim 500^{\circ}\text{C}$. Immediately after deposition, the sample was kept at the same temperature for an additional minute. The surface was then characterized using both LEED and Auger spectroscopy to assess the surface structure and the amount of deposited material.

Density functional theory (DFT) calculations were performed using the Quantum ESPRESSO code\cite{QE-2009,QE-2017} with ultrasoft pseudopotentials including spin-orbit coupling.\cite{DalCorso_2014} The surface was modeled by a Si(111) slab formed of four buckled layers, with the bottom surface passivated by hydrogen atoms to saturate dangling bonds.\cite{Higashi1990,Landemark1991} The two bottom Si layers were kept fixed at the experimental lattice constant, while the top two Si layers, the Bi atoms, and B dopants were allowed to relax. The energy cutoffs adopted in all calculations were 50 Ry for the wavefunctions and 450 Ry for the density. Structural relaxation was performed without including spin-orbit coupling and stopped when all force components were smaller than $10^{-3}$ Ry/Bohr. For both the structural optimization and the density calculations, the Brillouin zone was sampled with a $6\times6\times1$ shifted grid. The PBEsol exchange-correlation potential\cite{PBEsol} was used, and van der Waals interactions were included through the Grimme-D2 scheme.\cite{Grimme_2006}

\begin{figure}[!]
    \centering
    \includegraphics[width=8.5cm]{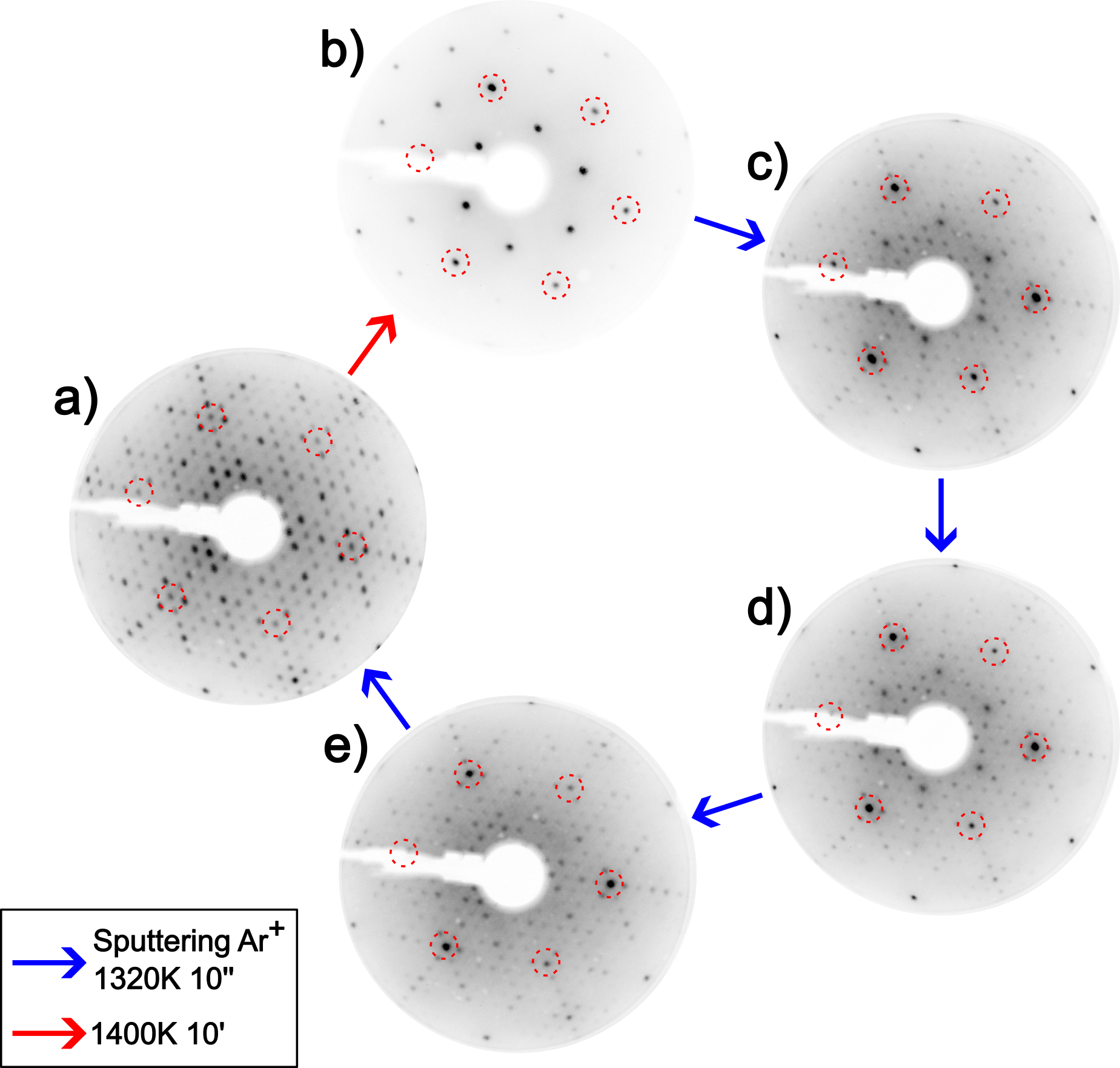}
    \caption{LEED sequence of the preparation of our samples, all taken at $75\text{eV}$. a) Clean Si(111) with $7\times7$ surface reconstruction. b) After 10' annealing at 1400K (red arrow), B-segregated Si(111) with $(\sqrt{3}\times\sqrt{3})R30^{\circ}$ pattern. c) After cycles of Ar$^+$ sputtering and 10" annealing at 1320K (blue arrow), both $7\times7$ and $\sqrt{3}\times\sqrt{3}R30^{\circ}$ are visible, indicating reduced B quantity. Continuing such cycles gives LEED d) then e), where B-induced spots got blurred. Eventually, the sample gets back to the initial state with a clean $7\times7$ Si(111) surface. Dashed red circles highlights the $1\times1$ diffraction spots of the Si(111) surface.}
    \label{Fig1}
\end{figure}

\section{Results}

\begin{figure}[!]
    \centering
    \includegraphics[width=8.5cm]{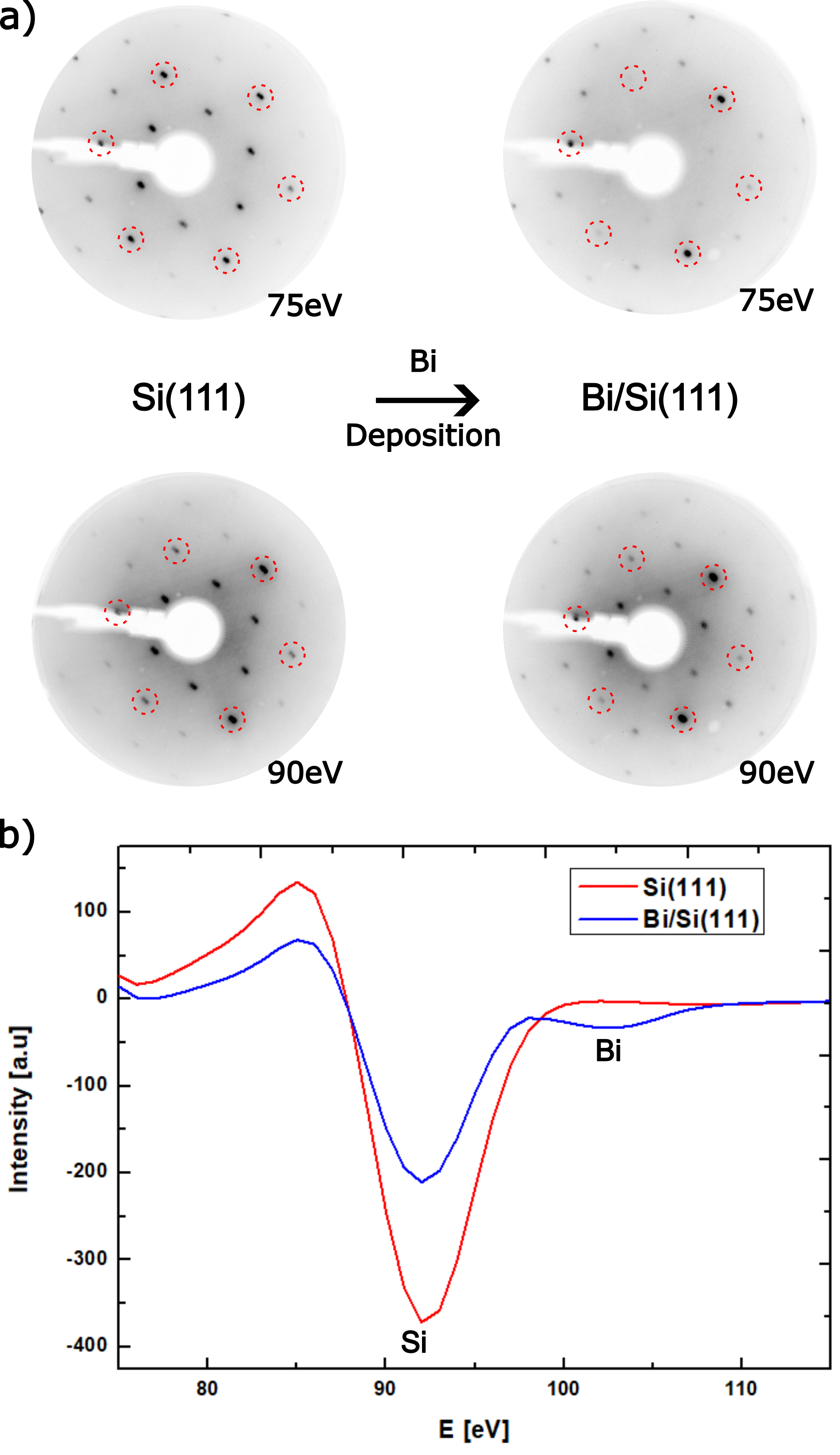}
    \caption{a) LEED pattern before (left) and after (right) Bi deposition on Si(111), in the case of B-segregated surface. Top images are taken at 75 eV and bottom ones at 90 eV. Dashed red circles highlights the $1\times1$ spots of Si(111). b) Auger spectroscopy spectra for B-segregated Si(111) (red) and Bi/Si(111) (blue). Si and Bi characteristic peaks 92 eV and 101 eV respectively) are indicated.}
    \label{Fig2}
\end{figure}

\begin{figure*}[!]
    \centering
    \includegraphics[width=17cm]{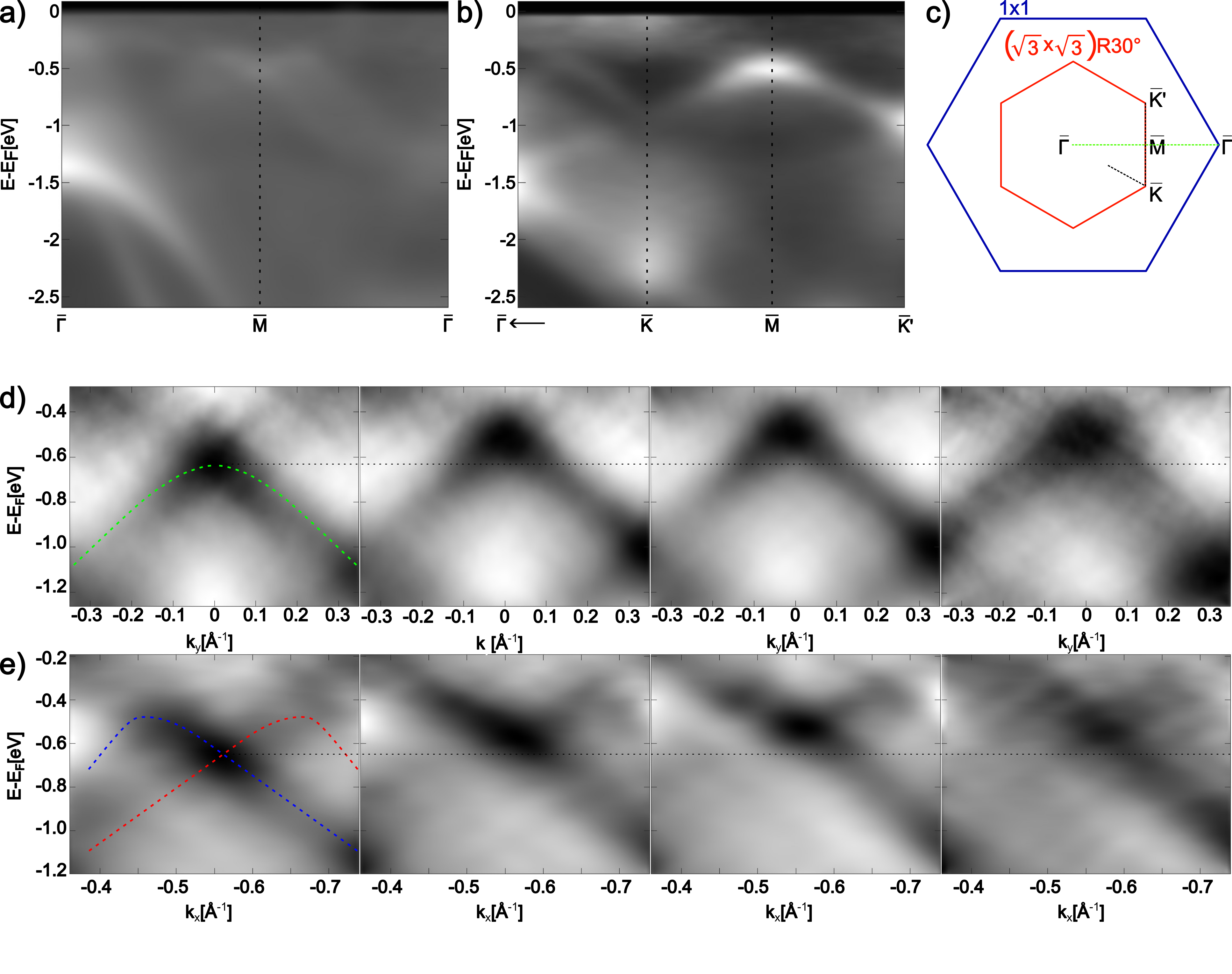}
    \caption{a),b) ARPES data along the $\overline{\Gamma}-\overline{\text{M}}-\overline{\Gamma}$ and $\overline{\Gamma}\leftarrow\overline{\text{K}}-\overline{\text{M}}-\overline{\text{K}}$ direction of the sample corresponding to LEED Fig\ref{Fig1}c). c) Schematic representation of the reciprocal space. Blue hexagon represents the first Brillouin Zone (BZ) of the $1\times1$ of Si and red hexagon the first BZ of the $(\sqrt{3}\times\sqrt{3})R30^{\circ}$. Dashed green and black lines mark the path of data represented in a) and b), respectively. 
    d),e) ARPES second derivative plots of our 4 samples, sorted by increasing B-doping, along the $\overline{\text{K}}-\overline{\text{M}}-\overline{\text{K}}$ and $\overline{\Gamma}-\overline{\text{M}}-\overline{\Gamma}$. Schemes of the bands in both directions are depicted on the left images, and dotted lines align with Rashba crossing point are represented to help see the upward band shift.
 }
    \label{Fig3}
\end{figure*}

It has long been established that dopants in silicon can diffuse through the crystal, with temperature playing a key role in this process.\cite{Jones2008DiffusionIS} In the case of boron, high-temperature annealing drives segregation to the surface.\cite{Thibaudau_1994,Zotov_1996,Lyo_1989,Thibaudau_1989,Dumas_1988} B atoms migrate into the second Si layer, occupying the so-called $T_4$ sites, where one out of every three Si atoms is replaced.\cite{Lyo_1989,Thibaudau_1994}. This results in a $1/3$ ML coverage of boron atoms that induces a $(\sqrt{3}\times\sqrt{3})R30^{\circ}$ surface reconstruction with respect to Si(111). Such segregation produces an extremely high local doping level in the near-surface region, which can be progressively reduced by controlled cleaning.

We exploited this mechanism to reproducibly tune the boron content at the surface. Cycles of Ar$^+$ sputtering followed by short annealing at 1320 K (10 s) gradually removed segregated boron. Figure~\ref{Fig1}a shows the LEED pattern obtained after sufficient cycles, corresponding to the clean $7\times7$ reconstruction of pristine Si(111). From this state, segregation could be re-induced by annealing at higher temperature and longer time. Figure~\ref{Fig1}b shows the sharp $(\sqrt{3}\times\sqrt{3})R30^{\circ}$ pattern that appears after annealing at 1400 K for 10 min, indicative of a full $1/3$ ML of B at the surface. Intermediate stages, illustrated in Fig.~\ref{Fig1}c–e, show coexistence of the $7\times7$ and $(\sqrt{3}\times\sqrt{3})R30^{\circ}$ patterns, with the intensity of the B-related spots decreasing gradually until they eventually blur out. This demonstrates that the surface B concentration can be adjusted in a controlled and reproducible way, providing a tunable knob to vary the near-surface doping level.

After this calibration step, Bi was deposited on Si substrates with different controlled amounts of segregated boron, as described in the Methods section. The formation of the $\beta$-phase of Bi/Si(111)$\,$—the trimer phase at 1 ML coverage—was confirmed by combining LEED and Auger spectroscopy. Figure~\ref{Fig2}a shows LEED patterns before and after Bi deposition in the case of a fully B-segregated substrate. For the clean B-segregated Si(111), the $(\sqrt{3}\times\sqrt{3})R30^{\circ}$ reconstruction is clearly visible at both 75 eV and 90 eV electron energies (left panels). After Bi deposition, the surface reconstruction is only visible at 90 eV, while at 75 eV only the $1\times1$ spots remain (right panels). This diffraction behavior was observed independently of the amount of boron present, and is therefore characteristic of the $\beta$-phase Bi/Si(111) structure. 

The Auger spectra in Fig.~\ref{Fig2}b provide further confirmation. After Bi deposition, both Si and Bi peaks are visible, and the change in relative Si peak intensity before and after deposition is consistent with a Bi coverage of approximately $1.1 \,\text{ML}$, assuming pseudomorphic growth and an electron inelastic mean free path of 5.4 \AA. \cite{Tanuma2011} Taken together, the LEED and Auger data establish that the Bi layer forms in the expected $\beta$-phase on all substrates, irrespective of the initial boron concentration at the surface.

Figure~\ref{Fig3} presents the electronic properties of the $\beta$-phase Bi/Si(111) measured by ARPES. Panels (a) and (b) show band dispersions along the $\overline{\Gamma}-\overline{\text{M}}-\overline{\Gamma}$ and $\overline{\text{K}}-\overline{\text{M}}-\overline{\text{K}}$ directions, respectively, for the sample corresponding to the LEED pattern of Fig.~\ref{Fig1}c. The schematic Brillouin zones of Si(111) and of the $(\sqrt{3}\times\sqrt{3})R30^{\circ}$ reconstruction are shown in the inset. The characteristic features of the $\beta$-phase are clearly visible: Rashba-split bands at the M point of the reconstructed Brillouin zone with a Rashba parameter of the order of $2 \,\text{eV}\cdot$\AA, and a simple parabolic dispersion along the $\overline{\text{K}}-\overline{\text{M}}-\overline{\text{K}}$ direction. These observations are fully consistent with previous experimental studies\cite{Gierz2009,Frantzeskakis_2010,Berntsen_2018,Gruznev_2014}. Intense bulk Si light- and heavy-hole bands appear at the $\Gamma$ point of the Brillouin Zone, at $\sim-1.3\text{eV}$ below the $\text{F}_\text{L}$ (the energy position is due to orthogonal wavevector selection), while another, much weaker band appears at $\sim-0.2\text{eV}$ below $\text{F}_\text{L}$, positioned $\sim0.3\text{eV}$ above the Bi-induced Rashba bands at the M point of the reconstruction, in agreement with theoretical studies. \cite{Chi2021,Gruznev_2014} Importantly, the band shapes are essentially identical for all B concentrations, demonstrating that boron segregation does not alter the crystalline phase or the topology of the Bi electronic structure. A slight broadening of the bands is observed in the case of fully segregated B surfaces (Fig.~\ref{Fig1}b), which may reflect a higher level of disorder in the Bi layer.

The systematic effect of B segregation appears instead in the energy position of the Bi Rashba bands. Panels (c) and (d) of Fig.~\ref{Fig3} show second derivative ARPES maps for four different samples with increasing B content, arranged from left to right. Along both $\overline{\text{K}}-\overline{\text{M}}-\overline{\text{K}}$ (c) and $\overline{\Gamma}-\overline{\text{M}}-\overline{\Gamma}$ (d) directions, the characteristic Rashba “M-shaped” dispersions are clearly visible in all cases. To quantify the band shift we track the energy of the crossing point at M (dotted lines), which provides a reliable reference. In the case of a p-doped Si(111) substrate without B segregation (left panels), the Rashba crossing point is found at $\sim 620 \,\text{meV}$ below the Fermi level, in agreement with the values previously reported for Bi/Si(111) on n-doped substrates.\cite{Gierz2009,Frantzeskakis_2010,Berntsen_2018,Gruznev_2014} As the boron content in the subsurface layer increases, the Rashba bands shift systematically upward by up to $\sim 150\!-\!200 \,\text{meV}$. The shift of the band apex is slightly larger (about 200 meV), but the overall shape and Rashba splitting remain unaffected. 

\begin{figure*}[!]
    \centering
    \includegraphics[width=17cm]{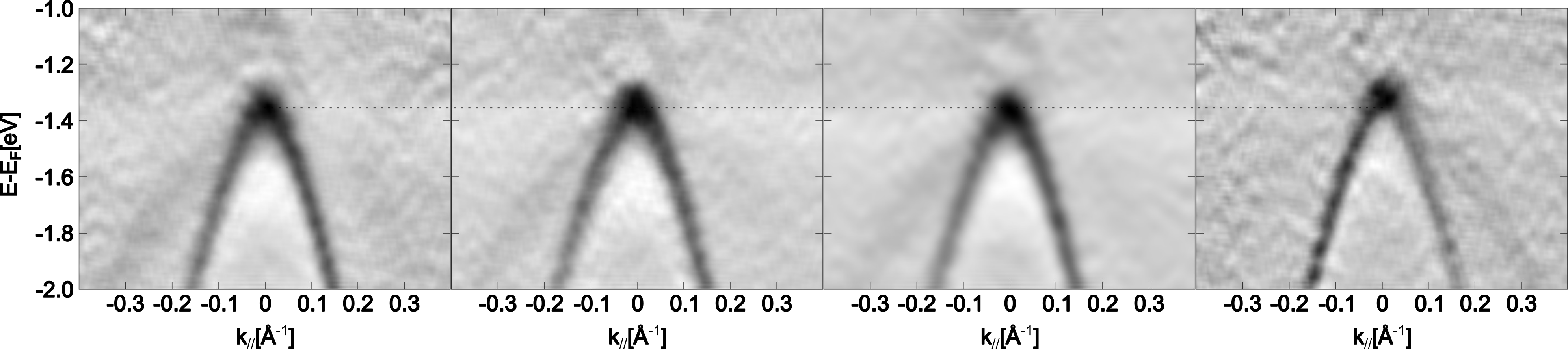}
    \caption{ARPES second derivative plots of our 4 samples at centered at the $\Gamma$ point, along the $\text{M}-\Gamma-\text{M}$ direction. The black dotted line highlights the evolution of the apex of the light holes band of Si(111).}
    \label{Fig4}
\end{figure*}

These measurements demonstrate that boron segregation provides an efficient knob to tune the energy position of the Bi-induced states, without altering their large Rashba parameter or the general topology of the electronic structure.

It is important to emphasize that the measured energy position of the Rashba crossing point in our p-doped substrates without B segregation is essentially the same as in previously studied n-doped substrates. This apparent insensitivity to the bulk doping type can be understood within the conventional band-bending framework of Si(111). Near the surface, the effective dopant concentration is statistically reduced, which tends to shift the Fermi level toward the middle of the gap. In addition, intrinsic surface states of Si(111) have been shown to strongly pin the Fermi level, typically at about 0.63 eV above the valence band maximum.\cite{Himpsel_1983,Zhang_2012,Kr_gener_2011} Both effects lead to a downward band bending for p-doped Si(111), such that the Fermi level at the surface becomes essentially fixed close to the same energy as in the n-doped case. As a result, the Bi-induced Rashba states, whose energy alignment is determined by their hybridization with the Si substrate, appear at a nearly identical binding energy of $\sim 600$ meV below the Fermi level, independent of whether the bulk Si is n- or p-doped. This explains why changing the bulk doping type alone does not suffice to shift the Bi bands, and highlights the need for a more local control of the surface potential, as achieved by boron segregation.

We now address the effect of boron segregation on the band alignment. In a simple band-bending scenario, the very high local p-doping induced by segregation (B occupying one out of three T$_4$ sites) would reduce the downward band bending at the surface and shift \emph{both} the Bi-derived and the Si-derived states upward by a comparable amount (i.e., an almost rigid shift of the near-surface band structure). Figure~\ref{Fig4} rules out this interpretation. While the Bi-induced Rashba-split bands at $\overline{\text{M}}$ move upward by about $150$–$200$\,meV as the B content increases (cf. Fig.~\ref{Fig3}c,d), the Si valence states at $\Gamma$ remain essentially fixed: the apex of the light-hole band shifts by at most $\sim 30$\,meV across the same series. 

Therefore, the simple band-bending scenario has to be discarded. The data instead point to a \emph{selective} modification of the relative alignment, in which the Bi states move with respect to Si. As we show below, DFT captures this behavior: the Bi Rashba states at $\overline{\text{M}}$ retain their Bi character and dispersion, whereas the Si states at $\Gamma$ undergo a redistribution of their wavefunction toward the surface in the presence of segregated B, altering the hybridization landscape without producing a significant electrostatic shift of the Si manifold.

\begin{figure*}
    \centering
    \includegraphics[width=0.98\linewidth]{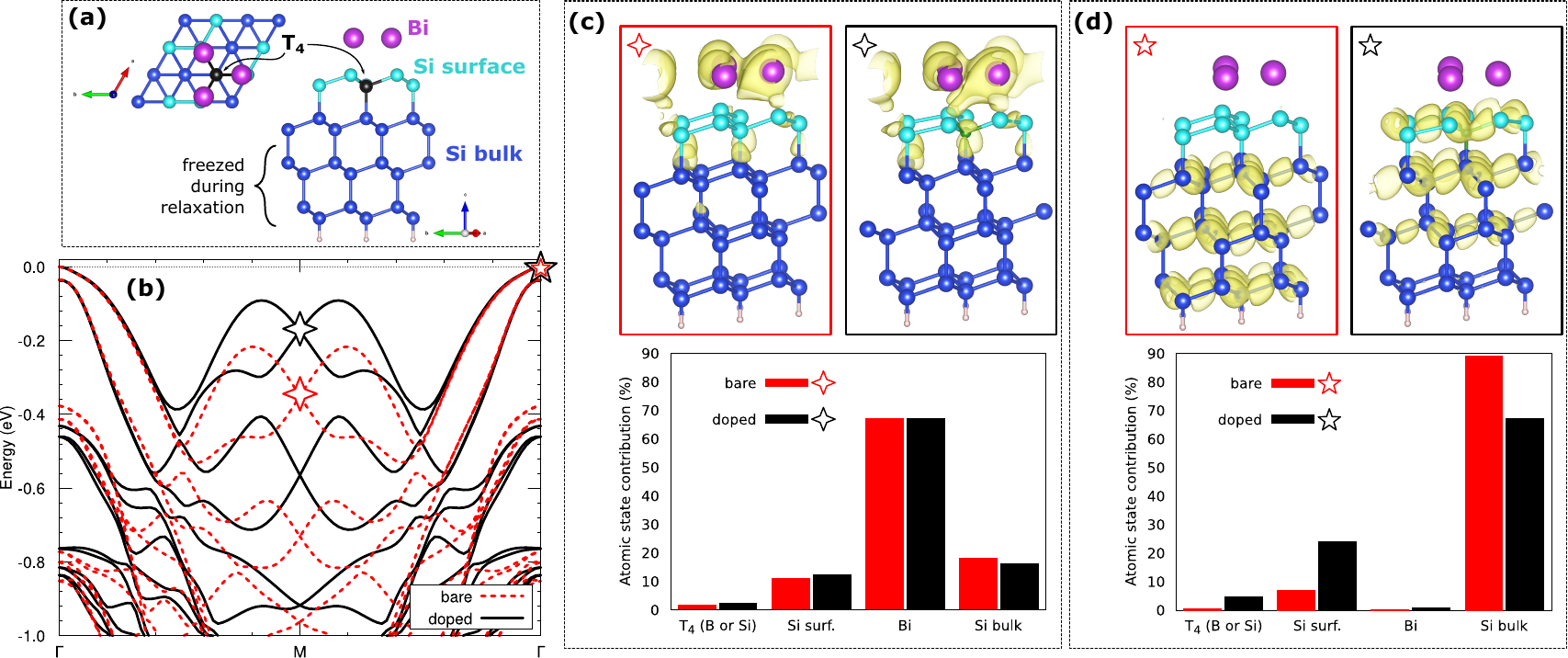}
    \caption{(a): Simulated structure from above and from the side. Different regions are highlighted with different colors: 3 Bi sites, 5 Si surface sites, 1 T$_4$ site (hosting either Si or B depending on the simulation), 18 Si bulk sites and 3 H sites.
    Details of the relaxation procedure are also provided.
    (b): Band structure of the bare (dashed red) and B-doped (solid black) simulations in the $\Gamma-M-\Gamma$ direction.
    States of interest are highlighted in $\Gamma$ (stars) and $M$ (diamonds) with the same color code as the bandplot.
    (c): 3D representation of the partial charge density $|\Psi|^2$ for the notable states at $M$ (on top) and histogram of its projection on atomic states belonging to different regions (bottom).  
    (d): Same as (c) but for the Si state at $\Gamma$.}
    \label{fig:dft-bands}
\end{figure*}

The model used in the DFT calculations, described in the method section, is sketched in Figure~\ref{fig:dft-bands}a.
We performed calculations in two different configurations: the bare one, with a Si in the $T_4$ site, and the doped one with a B in the $T_4$ site.

The corresponding band structures are reported in panel b of Figure~\ref{fig:dft-bands} in dashed red and solid black line respectively.
The electronic structure of the bare system is in good agreement with similar simulations\cite{Chi2021}.
We can focus on two remarkable features.
The band crossing at $M$ of the last two occupied bands, highlighted with a red diamond in the figure, is ascribed to Bi states\cite{Chi2021} and thus it can be put in relation with the measured Rashba crossing point. 
The second remarkable characteristic, which we highlighted with a red star, is the top valence at $\Gamma$ formed of parabolic bands coming essentially from Si.\cite{Chi2021}

Given the experimental evidence that Si bands do not move upon B segregation, the doped band structure has been aligned to the bare one in such a way that the Si state at $\Gamma$ (black star) is fixed.
With respect to the bare simulation, we observe an upward shift of the Rashba crossing point (black diamond) of 176 meV. The result is then consistent with the experiment and corroborates the selectivity of the method that can not be explained by a band bending.

To understand the origin of such a selective shift, we considered two possible effects: different variations of the potential acting on the Bi and Si states and a different modifications of their hybridization.

To evaluate the first effect, let $\Psi_j(\mathbf{r})$ be the wavefunction of the Rashba notable state ($j=M$) or the Si reference state ($j=\Gamma$).
Then, the expectation value of $V_\text{tot}(\mathbf{r})$ on these states reads
\begin{equation}
    \mathcal{V}_s = \int V_\text{tot}(\mathbf{r}) |\Psi_j(\mathbf{r})|^2 d\mathbf{r} \,,
\end{equation}
so that its variation when 
passing from the bare system ($s=b$) to the doped one ($s=d$) is $\Delta \mathcal{V} = \left(\mathcal{V}_d - \mathcal{V}_b\right) / \mathcal{V}_b$.

On both states, such a variation is of few \textperthousand, so it has a negligible influence on the shift we observed.

To evaluate the change in hybridization, we start by giving a closer look to the electronic wavefunctions of the notable states. In panels c and d of Figure~\ref{fig:dft-bands}, we report the 3D plot of the partial charge density --- that is $|\Psi_j(\mathbf{r})|^2$ --- in both the bare and the doped simulations.
Surprisingly, we observe that the Rashba state displays minor changes upon doping, while major modifications occur on the Si state at $\Gamma$.
In particular, in the bare calculation its wavefunction is mostly localized on the inner layers (bulk region), whereas it is pushed to the surface when B segregates at the surface.
A more quantitative view of this is given by the histograms below where
the atomic-like contributions to $|\Psi_j(\mathbf{r})|^2$ are reported for all the atoms according to the region or site they occupy.
While the Rashba state histogram is basically identical in the bare and doped systems, i.e. the hybridization of the state does not change, the contribution of surface Si atoms to the reference Si state increases upon doping at the expense of the bulk.

The conclusion we draw from this DFT analysis is that the segregation of B at the surface induces a change in the hybridization of the Si states whose energy is shifted down with respect to that of Bi states.
These, conversely, are almost unaffected by the surface segregation.
 
 In the experiment, this process results in an effective upward shift of the Rashba crossing point which moves closer to the Fermi energy because, as pointed out before, the position of the Si states with respect to the Fermi level is pinned.

\section{Conclusion}

In this work, we demonstrated that boron segregation at the surface of p-doped Si(111) provides an effective mean to tune the band alignment of the $\beta$-phase Bi/Si(111). By carefully controlling the segregation level through annealing and sputtering cycles, we achieved a reproducible upward shift of the Bi Rashba bands of up to $\sim 200$\,meV, while leaving the Si valence bands essentially unaffected. This selective behavior rules out a simple band-bending scenario and instead points to a modification of the local hybridization of Bi and Si states, as confirmed by our DFT analysis. 

Although the Rashba states remain below the Fermi level, their distance is nearly halved compared to the non-segregated case. This brings them significantly closer to accessibility for spintronic transport, and suggests that modest additional tuning, for example by electrostatic gating, could bring them to the Fermi level. More broadly, our results establish boron segregation and modulation doping as robust strategies for selective band engineering in two-dimensional metals on semiconductors. This approach could be extended to other systems where band alignment limits functionality, and may enable both device applications and the emergence of novel correlated or topological phases.

\bibliography{BiSiV1}

\end{document}